\documentclass{iau}
\usepackage{graphicx}
\usepackage{natbib}
\usepackage{amssymb,amsmath}

\title[Small asteroid system evolution] 
{Small asteroid system evolution}

\author[Seth A. Jacobson]   
{Seth A. Jacobson$^{1,2}$}

\affiliation{$^1$Laboratoire Lagrange, Observatoire de la C{\^o}te d'Azur, \\ 
Boulevard de l'Observatoire, CS 34229, F 06304 Nice Cedex 4, France \\
$^2$Bayerisches Geoinstitut, Universt{\"a}t Bayreuth, \\ 
D 95440 Bayreuth, Germany \\
email: {\tt seth.jacobson@oca.eu}}

\pubyear{2014}
\volume{xxx}  
\pagerange{1}
\jname{Complex Planetary Systems}
\editors{A. Lemaitre \& Z. Knezevic}
\begin{document}

\maketitle

\begin{abstract}
Observations with radar, photometric and direct imaging techniques have discovered that multiple asteroid systems can be divided clearly into a handful of different morphologies, and recently, the discovery of small unbound asteroid systems called asteroid pairs have revolutionized the study of small asteroid systems. Simultaneously, new theoretical advances have demonstrated that solar radiation dictates the evolution of small asteroids with strong implications for asteroid internal structure. We review our current understanding of how small asteroid systems evolve and point to the future.

\end{abstract}

\firstsection 
\section{Asteroid system morphologies}
Asteroid systems cannot be represented by a single distribution of characteristics, instead they are apportioned among a number of distinct morphologies defined by size, multiplicity, spin states and shapes. Asteroids are divided into small and large asteroids according to whether or not rotational acceleration from the YORP effect controls their spin states over other factors. We focus on small asteroids with YORP-driven evolution.
\subsection{Size-determined morphology}
\begin{figure}[t]
\begin{center}
\includegraphics[width=0.67\textwidth]{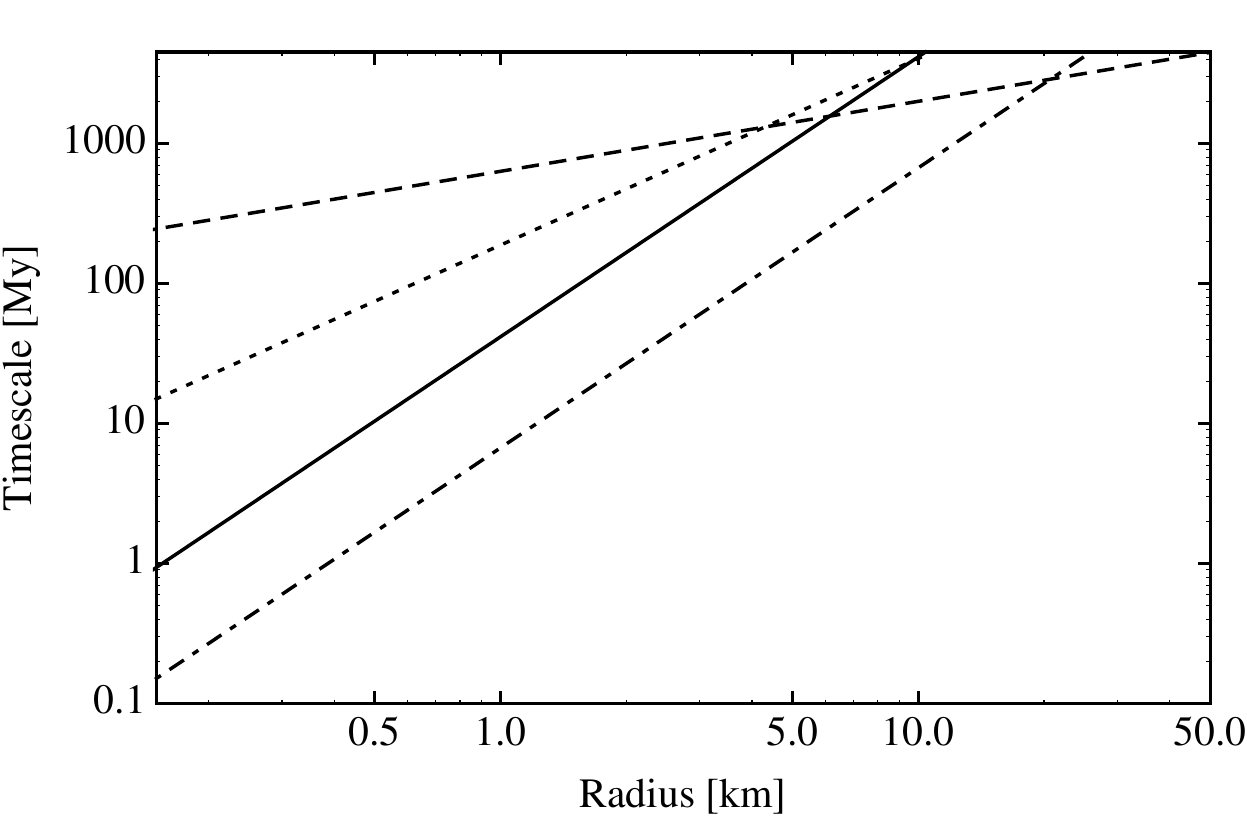}
\caption{Important small asteroid evolution timescales. Each line indicates the time in millions of years at which an asteroid of a radius $R$ given standard parameter choices will undergo the following processes: rotational acceleration to critical disruption spin rate $\omega_d$ from collisions in the main belt ($\tau_\text{rot}$, dashed line), catastrophic disruption due to collisions in the main belt ($\tau_\text{disr}$, dotted line), rotational acceleration to critical disruption spin rate $\omega_d$ from the YORP effect in the main belt ($\tau_\text{MBA}$, solid line), rotational acceleration to critical disruption spin rate $\omega_d$ from the YORP effect in the NEAs ($\tau_\text{NEA}$, dot-dashed line).}
\label{fig:Timescales}
\end{center}
\end{figure}

 The YORP effect is radiative torque due to asymmetrically emitted thermal photons~\citep{Rubincam:2000fg}. The timescale for an asteroid of radius $R$ to rotational accelerate from rest to the critical disruption spin rate is~\citep{Scheeres:2007kv}:
\begin{equation}
\tau_\text{YORP} = \frac{2 \pi \rho \omega_d R^2}{\bar{Y} H_\odot} 
\end{equation}
where $H_\odot = F_\odot / \left( a_\odot^2 \sqrt{1 - e_\odot^2} \right)$ is a term containing heliocentric orbit factors, $F_\odot \approx 10^{22}$ g cm s$^{-2}$ is the solar radiation constant, $a_\odot$ and $e_\odot$ are the semi-major axis and eccentricity of the asteroid system's heliocentric orbit, and the critical disruption spin rate $\omega_d = \sqrt{4 \pi \rho G / 3}$ is the spin rate at which the centrifugal accelerations match gravitational accelerations for a massless test particle resting on the surface of an asteroid of density $\rho$ (i.e. the spin rate that casts a test particle from the surface into orbit), $G$ is the gravitational constant, and, lastly, $\bar{Y}$ is the effective YORP coefficient, which takes into account small changes in obliquity and shape over the rotational acceleration of the asteroid. Furthermore, it is the sum of the normal YORP components~\citep{Rubincam:2000fg,Scheeres:2007kv} and the tangential YORP components~\citep{Golubov:2012kt}. Typical values for the effective YORP coefficient are $\bar{Y}\sim0.01$~\citep{Taylor:2007kp,Kaasalainen:2007hq}, and so characteristic timescales are $\tau_\text{NEA} \sim 6.7 \left(R^2 / 1\text{ km} \right)$ My for the near-Earth asteroid (NEA) population and $\tau_\text{MBA} \sim 41.6 \left(R^2 / 1\text{ km} \right)$ My for the main belt asteroid (MBA) population. These timescales are shown in Figure~\ref{fig:Timescales}. Amongst the NEAs, the YORP effect is more important than collisions~\citep{Bottke:1994vp} and planetary flybys~\citep{Rossi:2009kz}, because there are no NEAs significantly larger than a few tens of kilometers and those events are comparatively rare. Thus, the NEA population contains only small asteroids, however the YORP effect timescale can exceed the dynamical lifetime of a NEA, so observed NEAs have not necessarily undergone significant YORP effect evolution in NEA space.

The MBA population does contain both small and large asteroids. Collisions catastrophically disrupt asteroids of radius $R$ on a timescale $\tau_\text{disr} = 633 \left( R/1\text{ km} \right)^{1/2}$ My and deliver enough angular momentum to rotationally accelerate the asteroid significantly on a timescale $\tau_\text{rot} = 188 \left( R / 1\text{ km} \right)^{3/4}$ My~\citep{Farinella:1998ff}. These timescales are shown in Figure~\ref{fig:Timescales}, and it is clear that the division between small, YORP-driven asteroids and large asteroids is when the asteroid radius $R \sim 6$ km. This is confirmed by more sophisticated modeling~\citep{Jacobson:2014bi}. Furthermore, a division amongst the large asteroids at $R \sim 50$ km is apparent between those small enough to be significantly evolved by collisions during the current dynamical epoch of the main belt~\citep[e.g. $\sim4$ Gy for a Nice model explanation of the late heavy bombardment,][]{Marchi:2013du} and those even larger whose spin states on average reflect an earlier epoch, possibly primordial.
\begin{figure}[t]
\begin{center}
\includegraphics[width=0.8\textwidth]{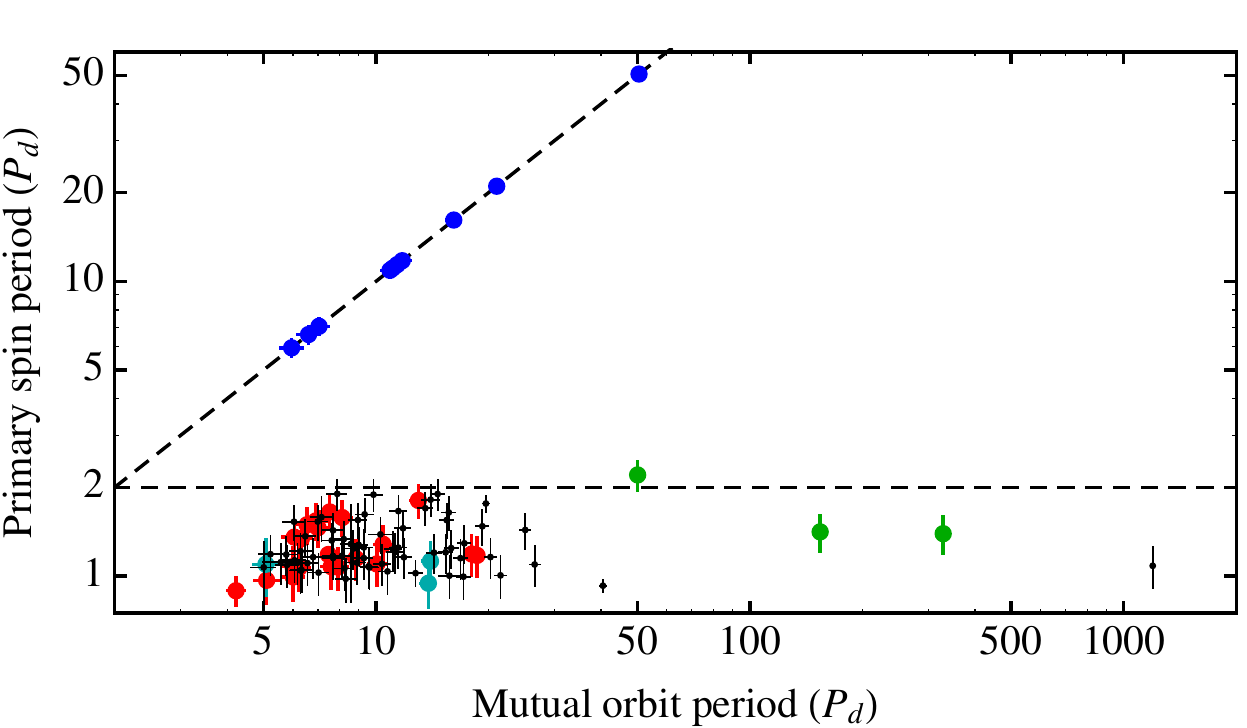}
\includegraphics[width=0.8\textwidth]{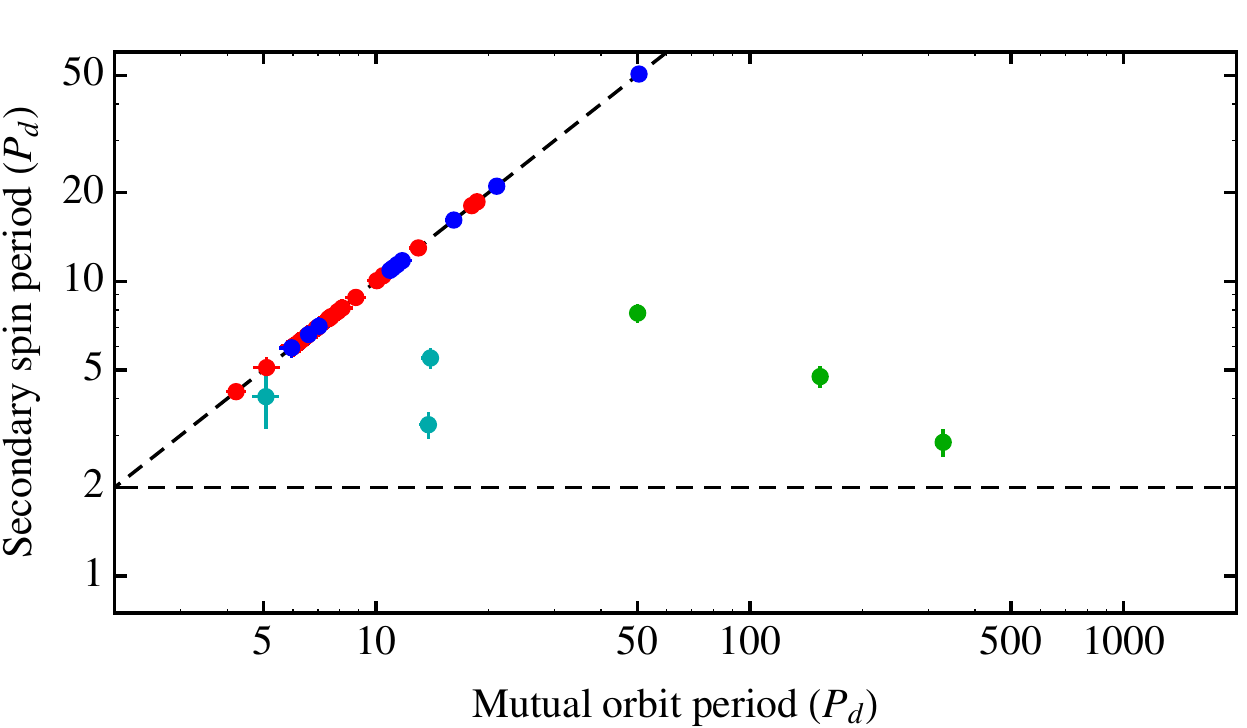}
\includegraphics[width=0.8\textwidth]{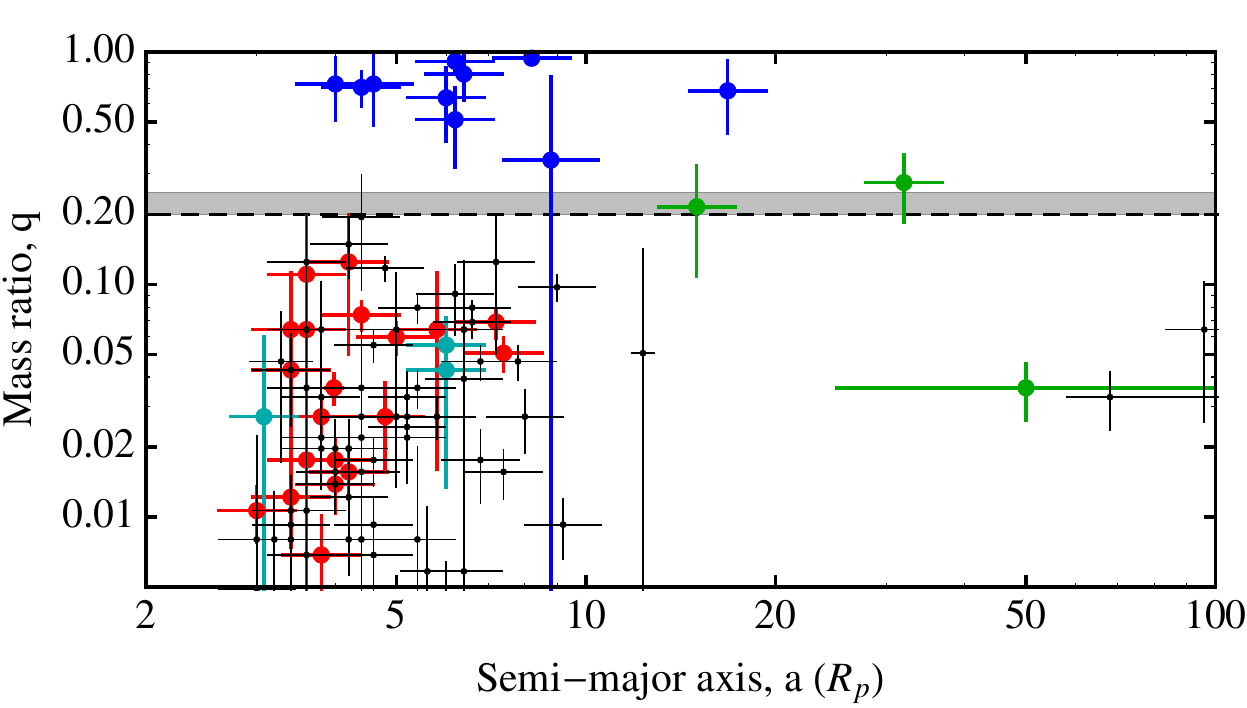}
\caption{Observational data for the small binary asteroid dataset from the September 2013 release of the Binary Asteroid Parameters dataset compiled according to~\citet{Pravec:2007fw} and updated by Petr Pravec and colleagues. Each binary is shown as a point with 3-$\sigma$ uncertainties shown (some uncertainties are very small). Small black points represent incompletely characterized binaries, i.e. no measured secondary spin period. Large points represent well-characterized binaries and their color indicates their morphology: blue is doubly synchronous, red is singly synchronous, cyan is tight asynchronous, and green is wide asynchronous. The upper graph shows the primary spin period $P_p$ measured in units of the critical disruption period $P_d = \sqrt{3 \pi / \rho G}$ as a function of the mutual orbit period $P_o$ measured in the same units. Any systems on the diagonal dashed line have a synchronicity between the primary spin periods and orbit periods. The horizontal dashed line at 2 $P_d$ is the approximate rapid rotator cutoff. The middle graph shows the secondary spin period $P_s$ as a function of the mutual orbit period $P_o$ measured in critical disruption periods $P_d$. Synchronous secondaries and the rapid rotator cutoff are again indicated by dashed lines. The bottom graph shows the mass ratio $q$ of each binary pair as a function of the semi-major axis $a$ measured in primary radii $R_p$. The dashed line indicates the special mass ratio of $q \approx 0.2$~\citep{Scheeres:2007io} and the gray region above it indicates some uncertainty up to $q \sim 0.25$ due to variations in shape~\citep{Taylor:2014by}.}
\label{fig:BinaryData}
\end{center}
\end{figure} 
\clearpage  
\subsection{Multiplicity-determined morphology}
Multiplicity is usually taken to refer to the number of bound members within an asteroid system, but it has been known since the beginning of the 20th century that unbound asteroids can be linked together. Unbound asteroid systems include the well known asteroid families formed from the debris from collisions and the recently discovered asteroid pairs~\citep{Vokrouhlicky:2008dt,Vokrouhlicky:2009kt,Pravec:2009hv}, which is a misnomer since asteroid `pairs' of more than two components exist~\citep{Vokrouhlicky:2009ja,Pravec:2013ud}. Color photometry and spectroscopic analysis have confirmed that asteroid pairs are compositionally related as well~\citep{Moskovitz:2012ga,Duddy:2012gb,Polishook:2014ub}. Unlike asteroid families, which are confirmed by examining the statistical significance of their number densities against the background asteroid population, asteroid pairs are confirmed through direct backwards numerical integration. This limits the timescale for which they can be discovered to $\sim$2 My because the uncertainty of the asteroid's position and velocity is chaotically lost due to perturbations from the planets and other asteroids~\citep{Vokrouhlicky:2008dt,Pravec:2009hv}. For discovered asteroid pairs, these dynamical ages are also likely the surface ages of the asteroid components and so they are a powerful tool to explore space weathering and other surface processes~\citep{Polishook:2014ub}. Unbound systems share a common ancestor, but are no longer interacting---a possible exception to this may be a feedback effect in large asteroid families, e.g. the increase in local number density increases the local collision rates.

Bound multiple asteroid systems are observed to contain two or three members, but there is no theoretical limit. Amongst the small multiple asteroid systems, there are distinct patterns related to the orbit period, spin periods and mass ratio as shown in Figure~\ref{fig:BinaryData}. Binary asteroid systems have a larger primary and a smaller secondary, and their mass ratio is the mass of the secondary divided by the mass of there primary. The most common observed small binary system is the singly synchronous, so named because the spin period of the secondary is synchronous with the orbit period. These binaries also have low mass ratios, which is defined as $q < 0.2$, tight semi-major axes, which is defined as $a \lesssim 8\ R_p$, and a primary with rapid rotation, which is defined as a period within a factor of two of the critical disruption period $P_d = 2 \pi / \omega_d = \sqrt{3 \pi / \rho G}$. The tight asynchronous binary systems also have rapidly rotating primaries and tight semi-major axes, but the secondary period is neither rapidly rotating nor synchronous with the orbit period. Wide asynchronous binary systems are very similar but naturally have larger semi-major axes. This distinction between tight and wide asynchronous binaries comes from comparing and contrasting with the singly synchronous binaries, respectively. All three morphologies are  low mass ratio with the exception of a few wide asynchronous binaries which skirt the $q\sim0.2$ cut-off. This is unlike the doubly synchronous binary asteroids, which all have high mass ratios $q > 0.2$, and take their names from the fact that all three periods in the system are synchronous. As shown in Figure~\ref{fig:BinaryData}, most of the non-classified binary systems are consistent with the singly synchronous or tight asynchronous morphologies.

There are also a few confirmed small asteroid triple systems. All have rapidly rotating primaries and low mass ratios between the smaller two members and the primary. Some of the inner ternary periods are confirmed synchronous. All of the outer ternary periods are asynchronous. Both (3749) Balam and (8306) Shoko are `paired' to other asteroids 2009 BR$_{60}$ and 2011 SR$_{158}$, respectively~\citep{Vokrouhlicky:2009ja,Pravec:2013ud}.
\subsection{Spin-determined morphology}
\begin{figure}[t]
\begin{center}
\includegraphics[width=\textwidth]{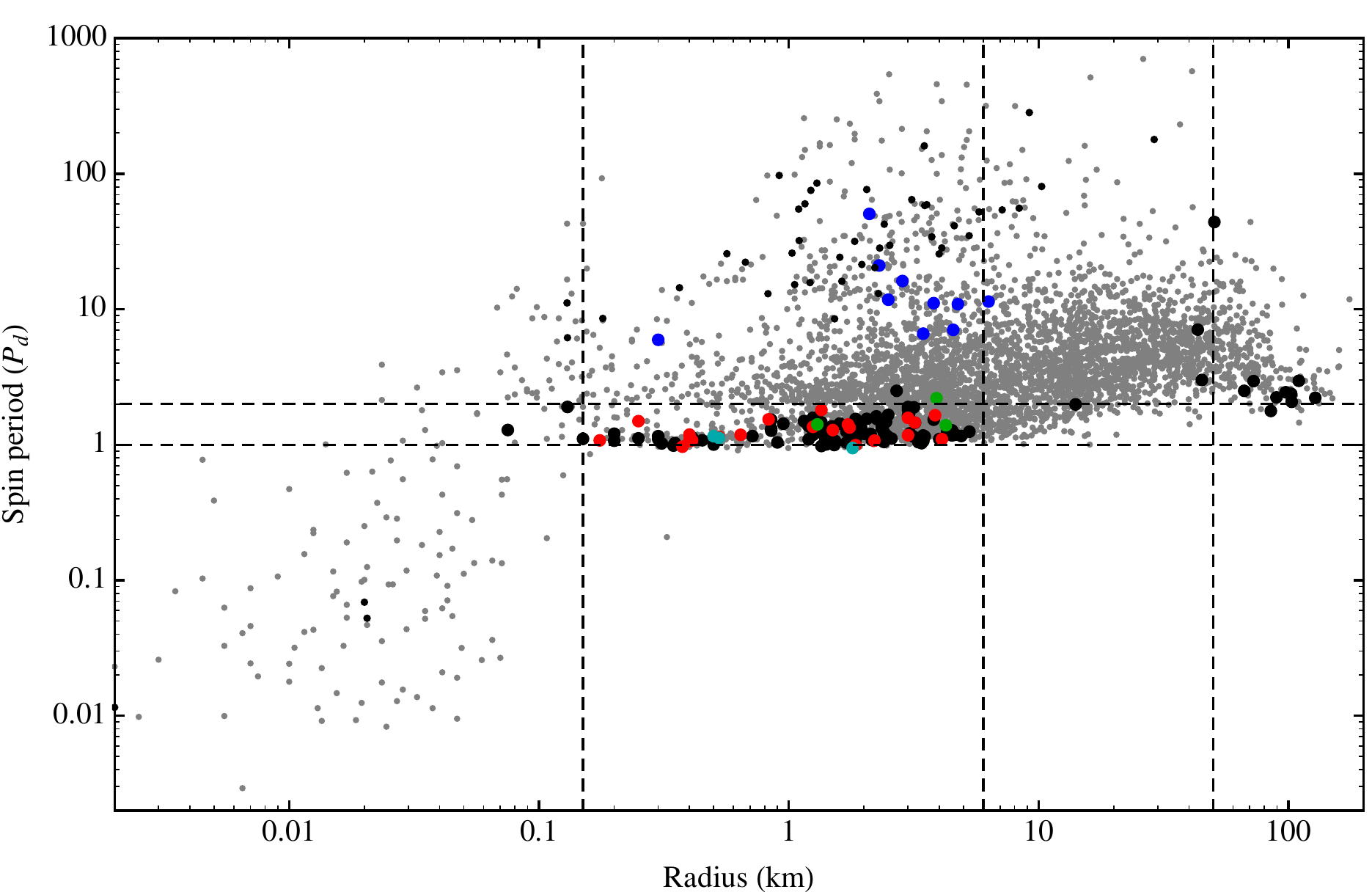}
\caption{The spin period of the NEAs and MBAs as a function of radius from the March 2013 release of the Asteroid Lightcurve Database compiled according to~\citet{Warner:2009ds} and updated by Brian Warner and colleagues. Single asteroids (or at least no detected companion) are shown as small gray points with the exception of the tumbling asteroids which are black. The primary spin periods and radii of multiple component asteroid systems are shown as larger points colored identically to Figure~\ref{fig:BinaryData}, and all known multiple component asteroid systems larger than 10 km are shown as large black points too. The spin period is measured in critical disruption periods $P_d = \sqrt{3 \pi / \rho G}$ assuming a bulk density $\rho = 2$ g cm$^{-3}$. The horizontal dashed lines indicate spin periods of 1 and 2 $P_d$, which brackets the rapid rotating population. The vertical dashed lines indicate radii of 6 and 50 km, which are identified as special in the text and Figure~\ref{fig:Timescales}.}
\label{fig:PeriodRadiusPlot}
\end{center}
\end{figure}
The distribution of spin periods amongst the asteroid population is a strong function of size as shown in Figure~\ref{fig:PeriodRadiusPlot}. Very small asteroids with radii less than 0.15 km are found at very high rotation rates well in excess of the rapid rotator population. It is difficult to observe the rotation of very small asteroids if the period is slow, since they are only observable for very short epochs. Small asteroids with radii between 0.15 km and 6 km, asteroids larger than this are not significantly torqued by the YORP effect (Figure~\ref{fig:Timescales}), have three distinguishing characteristics due to the YORP effect: rapid rotating single asteroids are common~\citep{Pravec:2000dr},  binary asteroids are common and mostly amongst the rapid rotator population~\citep{Pravec:2007fw}, and tumbling asteroids are common~\citep{Pravec:2005jf}. Amongst large asteroids with radii between 6 and 50 km, there is a dearth of rapid rotators, binary asteroids, and tumbling asteroids. It is difficult to assess whether this absence is real or an artifact, but since so many binaries have been discovered amongst larger and small systems, it is likely real unless there is a pathologic preference for a binary morphology that is difficult to detect such as extremely low mass ratios or wide semi-major axes. The dearth of tumblers is not as extreme as the dearth of rapid rotators. The very large asteroids with radii greater than 50 km do have binaries. The primaries are not rapid rotators but are amongst the fast rotators of this population.

While this work focuses on small asteroid evolution, it is worth noting that large multiple asteroids systems do appear very consistent with the formation of satellites in large impacts~\citep{Durda:2004en}. The very large primary asteroids with rotation periods near a few $P_d$ are consistent with the smashed target satellite formation mechanism (SMATS)~\citep{Durda:2004en} and a few wide asynchronous binaries (317 Roxanne and 1717 Arlon) are consistent with the escaping ejected binaries (EEB)~\citep{Durda:2010vz,Jacobson:2014hp}. These very large multiple asteroid system likely date to prior to the current collision environment ($>4 $ Gy) from timescale arguments (See Figure~\ref{fig:Timescales}). The dearth of binaries in the large asteroid morphology can be explained by considering that the YORP effect is not significant and that given the typically high velocities of collisions in the main belt, impacts on asteroids in this size range are too violent to create binaries and too frequent for systems created prior to the current collision environment to survive ($<4 $ Gy).
\subsection{Shape-determined morphology}
Lastly, it is worth noting that asteroids are not drawn from just a spectrum of oblate to prolate potato shapes, but also represent bi-modal, necked or contact binary shape distributions and diamond-shape distributions. Bi-modal and necked shapes range in appearance from two distinct asteroids resting on one another to only vaguely having a `head' and `body'. Collectively, they are often referred to as contact binaries. Diamond-shaped asteroids are often the primaries of binary asteroids, but single asteroids can also have this shape. This shape is characterized by an equatorial bulge, nearly oblate, shallow slopes near the equator and steep slopes near the poles.
\section{Rotational fission hypothesis}
\citet{Margot:2002fe} rejected formation of small binary asteroids from a sub-catastrophic impact, capture from a three-body interaction in the NEAs or MBAs, or capture after a catastrophic impact. None of these mechanisms systematically produce rapidly rotating primaries or the generally tight orbits observed amongst the binary populations. They concluded that rotational fission could explain these observations.

The rotational fission hypothesis posits that a parent asteroid can be torqued to a rotation rate so great that the centrifugal accelerations overpower the gravitational accelerations holding a rubble pile, i.e. strengthless, asteroid together~\citep{Weidenschilling:1980gz}. This rotational fission hypothesis is consistent with the most common singly synchronous binaries having super-critical amounts of angular momentum~\citep{Pravec:2007fw}. Planetary flyby-induced rotational fission was found to be insufficient to explain the existence of main belt and even near-Earth binaries~\citep{Walsh:2008gx}, but the YORP effect can rotationally accelerate asteroids to the critical disruption period, where rotational fission occurs~\citep{Scheeres:2007io,Walsh:2008gk,Sanchez:2012hz}. YORP-induced rotational fission is naturally limited to only sizes where the YORP effect controls the spin state, so any binaries larger than $R\sim6$ km are unlikely to be formed from this mechanism. Large binaries do not have rapidly rotating primaries or tight semi-major axes, and are consistent with the SMATS formation mechanism. It is likely that very small asteroids have cohesive or physical strength, and so in these cases the centrifugal accelerations must overcome these additional forces in order for the asteroid to fission~\citep{Pravec:2000dr,Sanchez:2014ir}. The rotational energy at the time of the rotational fission event determines the possible orbit and spin configuration immediately after the disruption~\citep{Scheeres:2007io}. If the asteroid breaks into two massive components, then the spin period necessary for disruption is much lower. If the mass ratio between the two components is nearly equal, then the binary system after the fission event cannot find an escape trajectory. If the mass ratio is very small, then the binary system can find an escape trajectory. For two spheres, the transition mass ratio between these two regimes is $q \sim 0.2$, which matches the cut-off between doubly synchronous binaries and the rest.

\begin{figure}[t]
\begin{center}
\includegraphics[width=\textwidth]{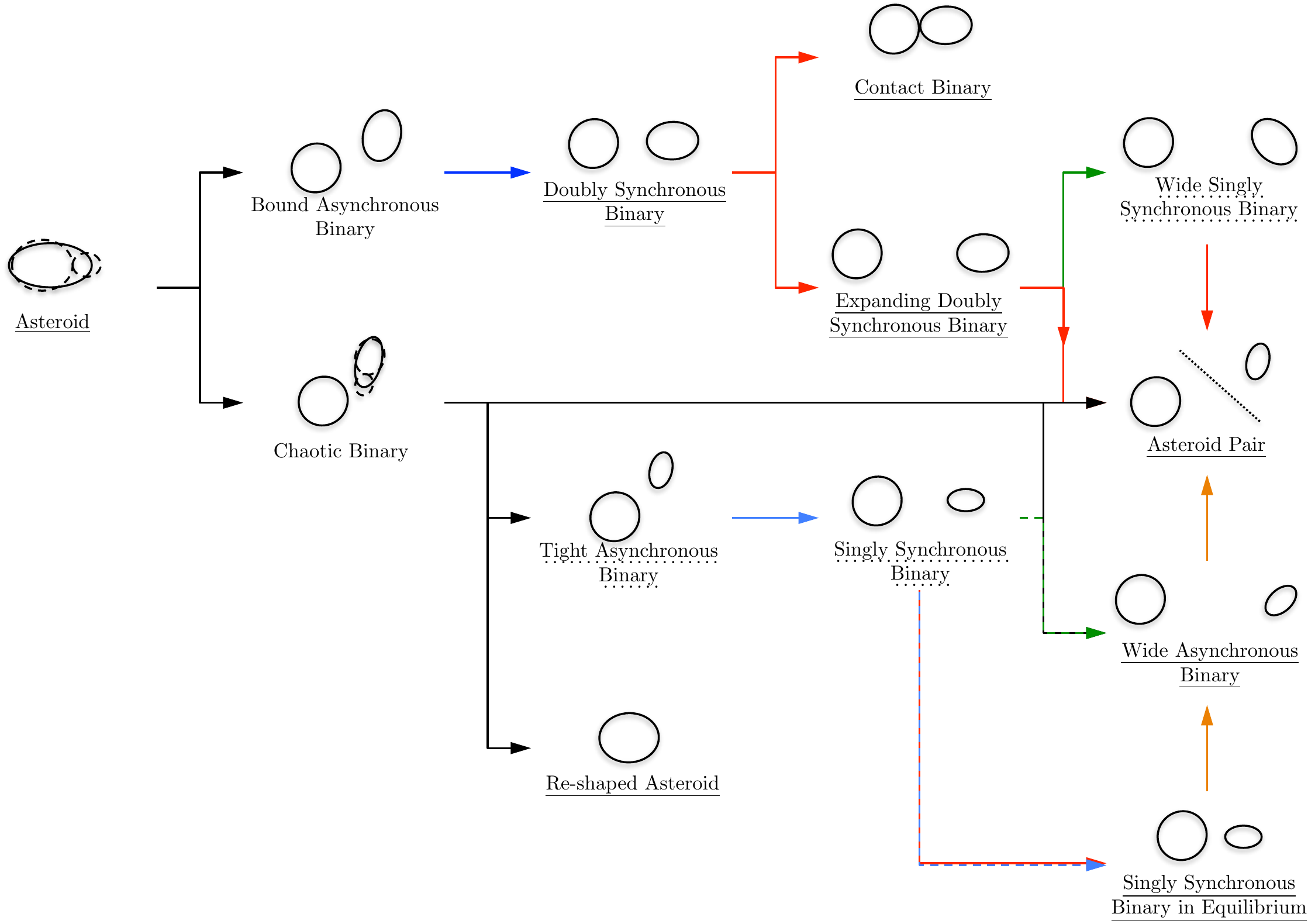}
\caption{A flowchart describing the evolution of an asteroid system after rotational fission from a single asteroid. All motion along the flowchart is left to right. Each state is described by a cartoon and labeled below. The typical timescale for an asteroid system to occupy that state is indicated by the underline: a solid underline indicates that the state persists until an external impulsive event such as a collision or planetary flyby or until a component undergoes YORP-induced rotational fission, a dashed underline indicates that the state is dynamically evolving according to a slow ($\sim10^{4}$--$10^7$ years), regular process such as tides, YORP and/or BYORP effect, and no underline indicates that the state is dynamically evolving on a fast ($\ll10^{4}$ years), irregular process such as strong spin-orbit coupling between the higher order gravity moments of the asteroid system components. The color of the arrow indicates the dominant evolutionary process: black indicates evolution dominated by gravitational interactions, blue indicates tides, red indicates BYORP effect, green indicates BYORP-driven adiabatic invariance de-synchronization, and yellow indicates a planetary flyby.}
\label{fig:flowchart}
\end{center}
\end{figure}
The evolution of small binary asteroid systems after a rotational fission event was mostly determined in~\citet{Jacobson:2011eq}, but there have been updates. The basic evolutionary flow is shown in Figure~\ref{fig:flowchart}. Particularly, we highlight alternative formation mechanisms for different morphologies including binaries with the primary synchronous, asteroid pairs that do not follow the mass ratio--primary period relationship in~\citet{Pravec:2010kt}, and triple systems.
\subsection{Low mass ratio track---chaotic binaries and asteroid pairs}
After a low mass ratio fission event, the resulting components enter into a chaotic orbit~\citep{Jacobson:2011eq}. Chaotic because the outcomes can only be described probabilistically. The fission process is likely very dusty~\citep{Polishook:2014ul}, although the dust will have nearly no relative velocity once it escapes the system. The first observations of an asteroid undergoing rotational fission are possibly being made today~\citep{Jewitt:2010bs,Jewitt:2014gp,Jewitt:2014wy}, although alternative explanations such as collisions and volatile-driven mass loss exist. Multiple event structures seen in this data are more likely due to repeated pericenter passages of the mutual orbit than repeated YORP-driven accelerations of an asteroid (since the YORP timescale is 100 years at best). During these pericenter passages angular momentum and energy are transferred between three reservoirs: the spin states of both components and the mutual orbit. These energy exchanges disrupt most systems on 1000 day timescales, and these disrupted systems are observationally consistent with the asteroid pair population---there is a predicted relationship between the mass ratio and the spin rate of the larger pair member~\citep{Pravec:2010kt}, and they have identical colors and spectrum~\citep{Moskovitz:2012ga,Polishook:2014ub}. Sometimes, they can find stable orbits. Often, the secondary is torqued so much by the spin-orbit coupling during pericenter passage that it undergoes rotational fission itself, i.e. secondary fission. Surface shedding produces the multiple event structures observed by~\citet{Jewitt:2010bs,Jewitt:2014gp,Jewitt:2014wy}, but secondary fission produces triple systems. A lot can happen to these triple systems. They may stabilize, a component may collide with the primary, or a component may escape the system.  Material leaving the secondary during secondary fission and impacting the primary is likely to collect on the equator creating the equatorial bulge and the ubiquitous diamond shape. Secondary fission is a process that can occur many times.
\subsection{Low mass ratio track---stable binary evolution}
Low mass ratio binary systems stabilize with semi-major axes between 2 and 17 primary radii $R_p$~\citep{Jacobson:2011eq}; the larger the stable orbit, the more eccentric the orbit is and the largest in numerical simulations were 17 $R_p$, which is a possible explanation for tight and wide binaries. Theoretically, stable binaries could be formed out to about 34 $R_p$, but these would be very eccentric~\citep{Jacobson:2014hp}. A simple analysis of eccentricity damping due to tides suggests that any mutual orbits larger than 10 $R_p$ are unlikely to damp, and so are still eccentric today. This may explain why some low mass ratio binaries have mild eccentricities. 

The fastest tidal process according to theory is synchronization of the secondary~\citep{Goldreich:2009ii}, and this is very consistent with the majority of tight binary systems being singly synchronous. Tight asynchronous binaries are likely younger than the average singly synchronous binary and the YORP torque on their smaller component is likely in the opposite direction of the synchronization tides. The combination of these two effects can explain the presence of a small population of tight asynchronous binary asteroids amongst a mostly singly synchronous population. Wide binary systems are asynchronous because synchronization tides fall off as the separation distance to the sixth power, and so are incredibly weak at these distances. Tides on the primary are very weak due to the low mass ratio compared to the YORP effect. Since the YORP effect likely has a positive bias from the tangential YORP component~\citep{Golubov:2012kt}, the primary maintains a spin period near the critical disruption period. If the primary accelerates enough, it could undergo rotational fission again and create a ternary system. This evolution has yet to be modeled in detail. However, spin-up in the presence of a tightly orbiting satellite is significantly different than companionless spin-up. Unlike companionless spin-up, the secondary's gravitational potential is constantly moving through the primary. This produces tides, but it also can set-off mini-avalanches and mass shedding events that maintain and perfect the diamond-shape. This process may be similar to those proposed in~\citet{Harris:2009ea}. Companionless asteroids are more likely to evolve catastrophically as they try to relax to the Maclaurin-Jacobi ellipsoid shapes~\citep{Holsapple:2010fv}. This would argue that the diamond-shape is not a result of YORP-driven rotational acceleration alone as suggested by~\citet{Walsh:2008gk}, but a combination of YORP acceleration and the at least temporary presence of a companion. 
\subsection{Low mass ratio track---singly synchronous binary evolution}
Singly synchronous binaries evolve due to both tides and the BYORP effect. The BYORP effect is a radiative torque on the mutual orbit of any synchronous satellite~\citep{Cuk:2005hb,McMahon:2010by}. It operates much like the YORP effect, and can substantially change the orbit of small binary systems in only $\sim 10^5$ years~\citep{McMahon:2010jy}. Since the BYORP effect can contract or expand the mutual orbit, there are two possible evolutionary paths. First, if the BYORP effect contracts the mutual orbit, then the singly synchronous system can reach a tidal-BYORP equilibrium~\citep{Jacobson:2011hp}. In this equilibrium, the mutual orbit circularizes and then no longer evolves. If it is, then the tidal-BYORP equilibrium can be used to determine important geophysical parameters. This binary state is stable and lasts until either the primary undergoes rotational fission, a possibility mentioned above, or a collision or planetary flyby occurs. It is possible that a planetary flyby could de-synchronize the secondary and expand the mutual orbit. If the expansion is great enough, then tides cannot synchronize the secondary and the system becomes a wide asynchronous system. Likely though, planetary flybys disrupt the binary forming an asteroid pair~\citep{Fang:2012go}.

If the BYORP effect expands the mutual orbit, then it is working with the tides. These binary systems may expand all the way to the Hill radius and disrupt forming asteroid pairs~\citep{Jacobson:2011hp}. These asteroid pairs will not follow the mass ratio--spin period relationship discovered in~\citep{Pravec:2010kt}, since the primary can spin back up after the rotational fission event but before the system disruption. Alternatively, an adiabatic invariance between the orbital period and the libration amplitude of the secondary can de-synchronize the secondary, when the orbit is very large. This creates wide asynchronous binaries on large circular orbits~\citep{Jacobson:2014hp}. This is in contrast to other formation mechanisms such as planetary flybys, direct formation from rotational fission or the escaping ejecta binary mechanism which all predict moderate to extreme mutual orbit eccentricity. Wide asynchronous binaries do not evolve by either tides (too wide an orbit) or the BYORP effect (asynchronous), so they can only be destroyed by collision or planetary flyby.
\subsection{High mass ratio track}
The high mass ratio track after rotational fission is much simpler. Since the two components have nearly the same mass, the tidal de-spinning timescales are nearly the same~\citep{Jacobson:2011eq}. Once both members are tidally locked, the system is a doubly synchronous binary, which agrees exactly with observations that all doubly synchronous binaries have high mass ratios. The BYORP effect does evolve these binaries, but the BYORP torques on each binary are independent. Likely, the observed binaries represent the population that have opposing BYORP torques, and so the mutual orbit is very slowly evolving. If the BYORP torques drain angular momentum from the system, then the binary may collapse~\citep{Cuk:2007gr,Taylor:2014by} forming a contact binary, but the BYORP torques can also expand the orbit. If they do so then, it can reach the Hill radius and form an asteroid pair. In this case, the asteroid pairs would have a mass ratio greater than $q\sim 0.2$. If the system evolved directly to the Hill radius then both would be rotating slowly, if one were to de-synchronize due to the adiabatic invariance but the other continues to drive the mutual orbit to the Hill radius from the BYORP effect, then the asteroid pair might have a rapidly rotating member, a slowly rotating member, and a mass ratio greater than 0.2.

\bibliography{biblio}
\bibliographystyle{astron}

\end{document}